\documentclass[10pt,twocolumn]{article}

\usepackage{multicol}
\usepackage{graphicx}
\usepackage{amsmath}
\usepackage{lipsum}
\usepackage{color}
\usepackage{siunitx}
\usepackage{authblk}
\usepackage[margin=2cm]{geometry}

\title{Mechanical tuning of the evaporation rate of liquid on crossed fibers}
\author[1]{Fran\c{c}ois Boulogne}
\author[1,2]{Alban Sauret}
\author[3]{Beatrice Soh}
\author[1,4]{Emilie Dressaire}
\author[1]{Howard A. Stone}
\affil[1]{Department of Mechanical and Aerospace Engineering, Princeton University, Princeton, NJ 08544, USA}
\affil[2]{Surface du Verre et Interfaces, UMR 125 CNRS/Saint-Gobain, 93303 Aubervilliers,  France}
\affil[3]{Department of Chemical and Biological Engineering, Princeton University, Princeton, NJ 08544, USA}
\affil[4]{Department of Mechanical and Aerospace Engineering, New York University Polytechnic School of Engineering, Brooklyn, NY 11201, USA}
\date{\today}
\begin{document}

\twocolumn[
    \begin{@twocolumnfalse}
        \maketitle
        \begin{abstract}
            We investigate experimentally the drying of a small volume of perfectly wetting liquid on two crossed fibers.
            We characterize the drying dynamics for the three liquid morphologies that are encountered in this geometry: drop, column and a mixed morphology, in which a drop and a column coexist.
            For each morphology, we rationalize our findings with theoretical models that capture the drying kinetics.
            We find that the evaporation rate significantly depends upon the liquid morphology and that the drying of the liquid column is faster than the evaporation of the drop and the mixed morphology for a given liquid volume.
            Finally, we illustrate that shearing a network of fibers reduces the angle between them, changes the morphology toward the column state, and therefore enhances the drying rate of a volatile liquid deposited on it.
        \end{abstract}
    \end{@twocolumnfalse}
]

\section{Introduction}

Evaporation is ubiquitous in natural phenomena such as the transpiration in plants \cite{Shantz1922}, the perspiration of humans and animals, and the dynamics of sea spray \cite{Andreas1995}.
Evaporation also impacts many technological applications including aerosol drying, food dehydration, and coatings.
However, modeling of the drying dynamics requires a description of the geometry  of the liquid, which is generally related to the wetting properties of the substrate.

When a drop is in contact with a substrate, the evaporation dynamics strongly depends upon the wetting properties.
The simplest situation of a drying drop consists of isolated aerosol droplets or non-wetting liquids \cite{Reyssat2008,Anantharaju2009,Gelderblom2011} for which the evaporation rate is set by the radius of the droplet \cite{Maxwell1877,Langmuir1918}.
For wetting liquids, the effect of the contact line must be taken into account \cite{Cazabat2010,Plawsky2008,Wayner1999}.
Indeed, the contact line generates a stronger evaporation flux because of a ``tip effect'' \cite{Cachile2002,Poulard2005,Guena2006,Guena2007,Berteloot2008,Pham2010}.
Furthermore, other properties of the substrate affect the drying such as the deformation of soft substrates \cite{Lopes2012} or drop confinement between two plates \cite{Clement2004}.

So far, most experiments address the drying of droplets resting on a plane but the drying of a drop on other substrates with different geometries remains largely unexplored.
For example, the evaporation of a liquid volume between fibers has industrial applications but remains poorly explored.
Indeed, networks of fibers such as fabric, filters and paper are a collection of crossing fibers.
In particular, at the node of two fibers, the liquid can adopt three different equilibrium morphologies that have been studied only recently \cite{Sauret2014}.

Previous studies have investigated the morphology adopted by liquid on fibers.
To understand the general situation of a complex three-dimensional network of fibers \cite{Duprat2012}, the wetting has been studied in different fiber configurations.
When a liquid is coated on a single fiber \cite{Quere1988,Quere1989,Quere1999}, depending upon its rheological properties \cite{Boulogne2013a,Boulogne2013b}, the liquid can undergo a Rayleigh-Plateau instability, which results in a series of droplets \cite{Rayleigh1878}.
A liquid droplet placed on a single fiber can adopt a barrel shape, a clam-shell shape or detach as gravitational forces dominate capillary forces \cite{McHale1999, McHale2001,Lorenceau2004}.

Other studies have investigated geometries involving a pair of fibers.
Observations of the equilibrium morphology of a wetting liquid on two parallel fibers has revealed that two morphologies can be adopted by the liquid.
When the distance between two fibers is large, the liquid is in a drop morphology in contact with both fibers.
Below a critical distance between the two fibers, the fluid spreads into a column that spans the gap between the fibers \cite{Princen1970,Protiere2012}.
Consequently, the interfacial area per unit volume of liquid increases significantly when the liquid changes morphology from a drop to a column, which speeds up the evaporation of the liquid \cite{Duprat2013}.
More recently, the investigation of the more complex situation of crossed fibers has revealed a rich morphology diagram as illustrated in figure \ref{fig:setup} \cite{Sauret2014}.
Three morphologies can be obtained depending upon the volume of liquid and the tilt angle $\delta$ between the fibers.

In this paper, we investigate the drying of a completely wetting liquid on crossed touching fibers with different tilt angles.
We report experimental results for all morphologies adopted by the liquid on crossed fibers, \textit{i.e.} the drop, the column and the mixed morphology.
For each of the morphologies, we provide a model to describe the time evolution of the liquid shape during drying.
We demonstrate that the tilt angle controls the efficiency of liquid evaporation and apply our findings to the evaporation of liquid on networks of fibers.
The observation that the tilt angle between fibers influences the evaporation rate indicates that mechanical effects can affect drying of wet fibrous materials.

\section{Experimental section}

\begin{figure}
    \centering
    \includegraphics[width=0.92\linewidth]{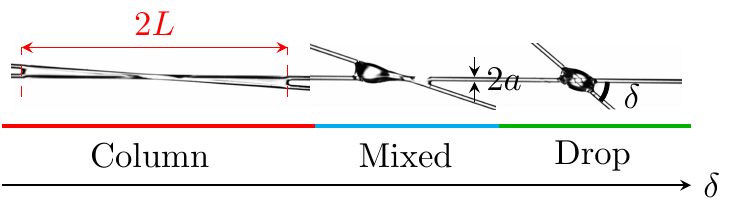}
    \caption{Pictures of the three possible morphologies for increasing tilt angle $\delta$ of the fibers: column (red), mixed (blue) and drop (green).
        The fiber diameter is $2\,a$ and the total column length is $2\,L$.
    }\label{fig:setup}
\end{figure}

Two nylon fibers of circular cross-section and of the same diameter $2a$ ($0.28$ or $0.30$ mm) are clamped on two horizontal rotating microcontrollers (PR01, Thorlabs).
The tilt angle $\delta$ (see figure \ref{fig:setup}) between the two fibers is adjusted using the microcontrollers.
We deposit a liquid drop of volume $V\in[0.5;\,4]\,\mu{\rm L}$ using a micropipette on the node of the two fibers.
Pictures are taken using Nikon cameras (D5100 and D7100) and $105$ mm macro objectives to observe both the side and the top views.
The tilt angle $\delta$ between the two fibers is measured directly from the top views by image processing with an uncertainty of $0.02^\circ$.
To ensure that the drying conditions are controlled, the entire setup is placed in an acrylic box to limit air convection.

For all of the experiments presented in this paper, we use completely wetting volatile silicon oils purchased from Sigma-Aldrich:
octamethyltrisiloxane (OMTS) with a kinematic viscosity $1$ mm$^2$/s, a density $\rho_\ell = 818$ kg/m$^3$ and a surface tension $\gamma=17.4$ mN/m,
decamethyltetrasiloxane (DMTS) with a kinematic viscosity $1.5$ mm$^2$/s, a density $\rho_\ell = 854$ kg/m$^3$ and a surface tension $\gamma=18.0$ mN/m.
These silicon oils are composed of a monodisperse oligomer that ensures a single drying kinetics.

The evaporation speed $V_E$ is obtained by recording the mass evolution $m(t)$ of a layer of liquid placed in a petri dish of radius $R_p=1.6$ cm and weighed using a high-precision scale (XS105, Mettler Toledo).
The evaporation speed can be defined as \cite{Duprat2013}
\begin{equation}V_E = \frac{1}{\rho_\ell \,S_p} \frac{\textrm{d}m(t)}{\textrm{d}t},\end{equation}
    where $S_p=\pi R_p^2$ is the surface area of the petri dish and $\rho_\ell$ the liquid density.
    The measured  evaporation speeds are $V_E = 3.9 \times 10^{-8}$ m/s and  $V_E = 3.6 \times 10^{-9}$ m/s for OMTS and DMTS, respectively.
    Note that it is consistent that the higher molecular weight (\textit{i.e.} higher viscosity) liquid evaporated more slowly.

    \section{Results}

    \subsection{Observations}

    \begin{figure}
        \centering
        \includegraphics[width=0.9\linewidth]{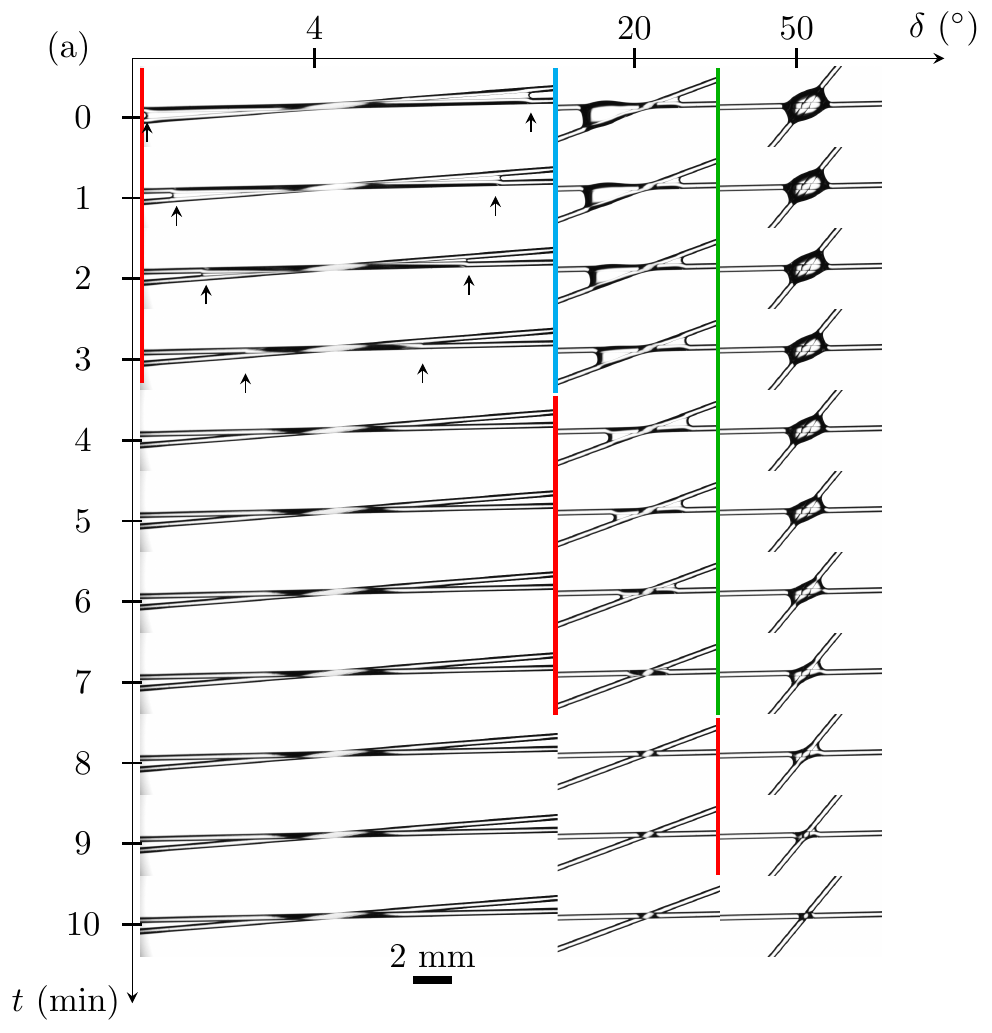}
        \includegraphics[width=0.9\linewidth]{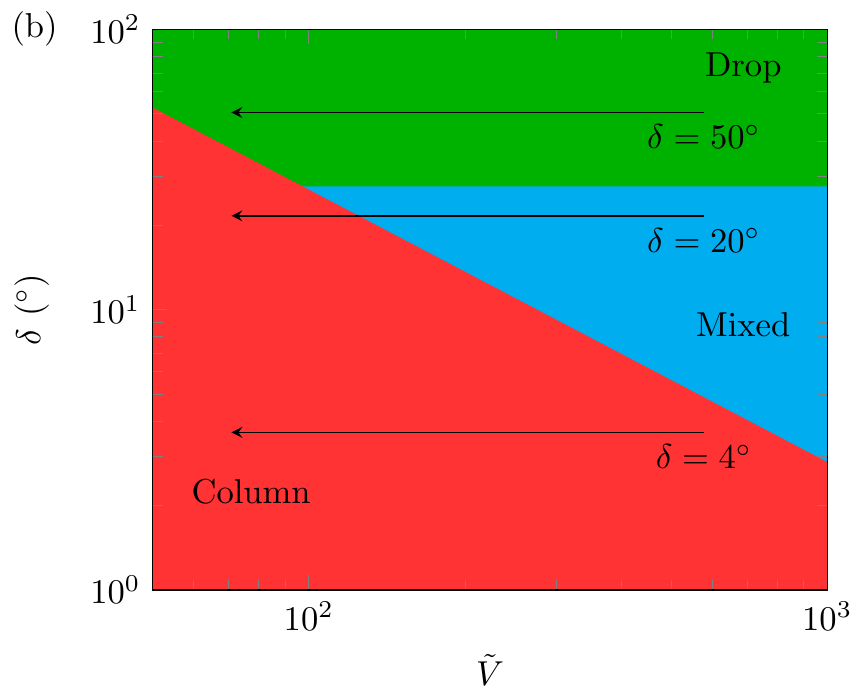}
        \caption{(a) Time evolution of a drying liquid at the node of two fibers for three different tilt angles $\delta$.
            The initial volume is $2$ $\mu\ell$ (OMTS) and the fiber diameter is $2\,a = 0.30$ mm.
            Black arrows indicate the position of the liquid column extremities.
            Colored lines on the left of each picture refer to the liquid morphology (see figure \ref{fig:setup}).
            (b) Phase diagram of the liquid morphology as a function of the tilt angle $\delta$ and the dimensionless volume $\tilde V = V / a^3$.
            Arrows indicate the evolution of the system in the phase diagram during the drying process for the three tilt angles.
        }\label{fig:dyn}
    \end{figure}

    The drying of liquid lying on crossed fibers, involves three different morphologies (figure \ref{fig:setup}) \cite{Sauret2014}.
    Indeed, as shown in figure \ref{fig:dyn}, the morphology adopted by the liquid depends upon the tilt angle $\delta$ and the drop volume $V$ (see supplementary information video) and changes in time during drying.
    In particular, as evaporation proceeds, the volume decreases and the new equilibrium morphology can be read on a morphology diagram for a given value of the tilt angle (see examples in figure \ref{fig:dyn}(b)).
    Therefore, we observe that liquids initially in the column morphology remain a column throughout the entire evaporation process while liquids initially in the mixed morphology switch to the column state.
    When the initial morphology of the liquid is a drop, the liquid can be assumed to remain a drop since the transition to the column state occurs at a very low volume.
    These states and transitions are summarized in figure \ref{fig:dyn}(b).

    \subsection{Theoretical background}

    Denoting $\rho_v$ the vapor concentration in the gas surrounding the liquid volume, Fick's second law is
    \begin{equation}
        \frac{\partial \rho_v}{\partial t} = D_{m} \nabla^2 \rho_v,\label{eq:fick}
    \end{equation}
    where $D_m$ is the diffusion coefficient of the vapor in the gas phase.
    Considering a characteristic lengthscale $\ell$ and an evaporation speed $V_E$, the P\'eclet number $\rm{Pe}$ \cite{Sultan2005} is the ratio of the diffusive time of the vapor in the gas phase $\tau_D = \ell^2/D_{m}$ to  the evaporation time $\tau_{E}=\ell/V_E$:
    \begin{equation}
        \textrm{Pe} = \frac{V_E \, \ell}{D_{m}}.
    \end{equation}
    The diffusion coefficient $D_m$ is derived from the kinetic theory of gases, $D_{m} \sim 10^{-6} $ m/s$^2$ \cite{Maxwell1867}.

    In our experiments, the characteristic length scale $\ell$ is the typical size of a drop, which is also the order of the capillary length $\kappa^{-1} = \sqrt{\gamma/(\rho_\ell g)} \simeq 1.5$ mm, where $g$ is the gravitational acceleration. The characteristic P\'eclet number is thus small, $\textrm{Pe} \ll 1$: the gradient of vapor in the gas surrounding the liquid is established much faster than the evaporation of the liquid.
    Consequently, the time derivative in equation (\ref{eq:fick}) is negligible.
    Therefore, the drying dynamics can be described by the Laplace equation:

    \begin{equation}
        \nabla^2 \rho_v = 0.
    \end{equation}
    We define the flux $j_0$ as the volume of liquid that evaporates per unit area and unit time, and write \cite{Cazabat2010}

    \begin{equation}
        \left.\frac{\partial \rho_v}{\partial r}\right|_{r=r_0} = - \frac{\rho_\ell\, j_0}{D_m}\label{eq:gradvap}
    \end{equation}
    where $r_0$ denotes the position of the interface.
    Assuming that $\rho_v(r=r_0)=\rho_{sat}$, the vapor density at saturation, equation (\ref{eq:gradvap}) provides the typical magnitude for the diffusive flux $j_0$

    \begin{equation}
        j_0 \simeq \frac{D}{\mathcal{L}}
    \end{equation}
    where $D = D_m \, \rho_{sat}/\rho_\ell$ and $\mathcal{L}$ is the length scale of the vapor concentration gradient.

    In the present paper, we assume that the flux $j_0$ is uniform over the air/liquid interface \textit{i.e.} we neglect the increase of the evaporation flux near contact lines.
    Thus, we can write that
    \begin{equation}
        \frac{\text{d} V}{\text{d}t}   \simeq - j_0   \mathcal{A} \simeq - \frac{D \mathcal{A}}{\mathcal{L}}\label{eq:master}
    \end{equation}
    where $V$ is the volume of liquid and $\mathcal{A}$ the total interfacial area.
    Equation (\ref{eq:master}) shows that the drying rate depends upon the morphology of the liquid through the interfacial area $\mathcal{A}$ and the lengthscale of the vapor concentration gradient $\mathcal{L}$, which is set by the liquid geometry.
    Next, we analyze different liquid morphologies and evaluate these two parameters, $\mathcal{A}$ and $\mathcal{L}$, to model the drying dynamics.

    \subsection{Drying of a droplet at the node of two crossed fibers}

    \begin{figure}
        \centering
        \includegraphics[width=0.9\linewidth]{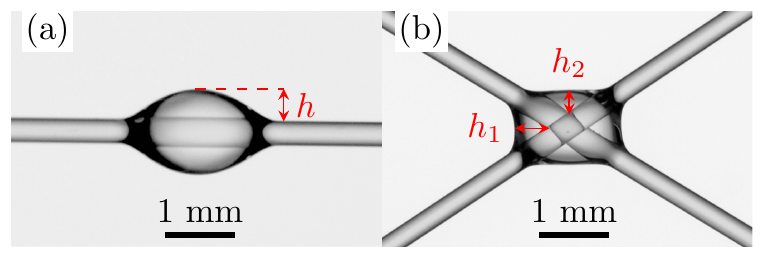}
        \includegraphics[width=0.9\linewidth]{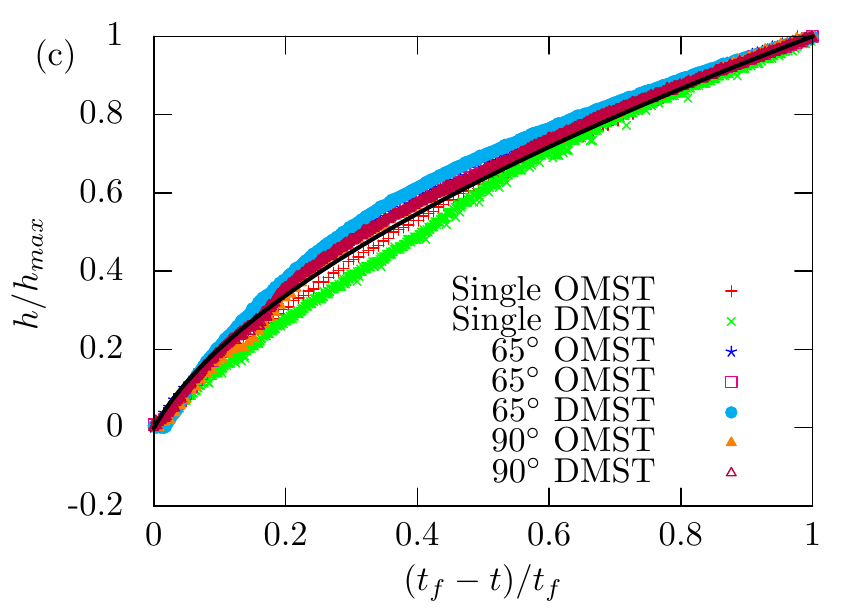}
        \caption{
            Definition of the fluid height, $h$, (a) for a single fiber and (b) for crossed fibers, $h = (h_1 + h_2)/2$.
            (b) Drying dynamics of drops in three configurations: single fiber and crossed fibers with tilt angles $\delta=65^\circ$ and $90^\circ$.
            The plot represents the evolution of the height $h/h_{max}$ with $(t_f - t)/t_f$, where $h_{max}$ is the initial height and $t_f$ is the total duration of the drying process determined experimentally.
            In each case, two liquids are used (OMTS and DMTS). The initial volume is $1$ $\mu\ell$ and the fiber diameter is $2\,a =0.30$ mm.
            The black line is the theoretical prediction given by equation (\ref{eq:eqdiff_drop}).
        }\label{fig:droplet}
    \end{figure}

    The drying dynamics of a spherical drop are set by the radius of the drop $R$, \textit{i.e.} $\mathcal{L} = R$ \cite{Cazabat2010}.
    Thus, the flux is $j_0 \simeq {D}/{R}$ and the drop radius decreases according to $R(t) = \sqrt{2D(t_f-t)}$ with $t_f$ being the total evaporation time.

    For a drop on a single fiber in an axisymmetric barrel shape \cite{McHale1999, McHale2001}, the drop is characterized by its length $L$ and its height $h$ (figure \ref{fig:droplet}(a)).
    The height $h$ is defined as the distance between the drop apex and the fiber surface.
    Assuming that the typical lengthscale of the vapor concentration gradient  is $L/2$ \cite{Duprat2013}, the evolution of the drop height is described by \cite{Carroll1976}:

    \begin{equation}
        \frac{\text{d} h}{\text{d} t} = - \frac{2 D \mathcal{A}}{ L (\text{d} V / \text{d} h)}.\label{eq:eqdiff_drop}
    \end{equation}
    The interfacial area $\mathcal{A}$ and the volume $V$ can be obtained analytically from Carroll's derivations \cite{Carroll1976,Duprat2013}.

    In the case of crossed fibers, the definition of the fluid height needs to be adapted as illustrated in figure \ref{fig:droplet}(b).
    The distance is equal to the average along the two bi-sectors.
    It is noteworthy that the lengthscales of the droplet on single and crossed fibers are similar.
    In figure \ref{fig:droplet}(c), we report the evolution of the drop height $h$ rescaled by its initial value $h_{max}$ as a function of $(t_f - t) / t_f$.
    The total duration of the drying process $t_f$ is determined experimentally from the direct observation that no liquid remains on the fibers.
    The numerical solution of equation (\ref{eq:eqdiff_drop}) is calculated with an iterative method as explained in the appendix of reference \cite{Duprat2013}.
    In both single and crossed fibers configurations, the drying dynamics follow a trend that is in a good agreement with the numerical solution.

    In addition, we observe good agreement between experiment and theory for the two liquids with different volatilities in each configuration as reported in figure \ref{fig:droplet}.
    Thus, the results validate that the vapor concentration gradient, and thus the dynamics, is set by the size of the droplet.
    Consequently, the evaporation of a drop at the node of two crossed fibers can be fairly well described by a model similar to the one developed for a drop on a single fiber.

    \subsection{Drying of liquid columns}\label{sec:column}
    \begin{figure}
        \centering
        \includegraphics[width=0.9\linewidth]{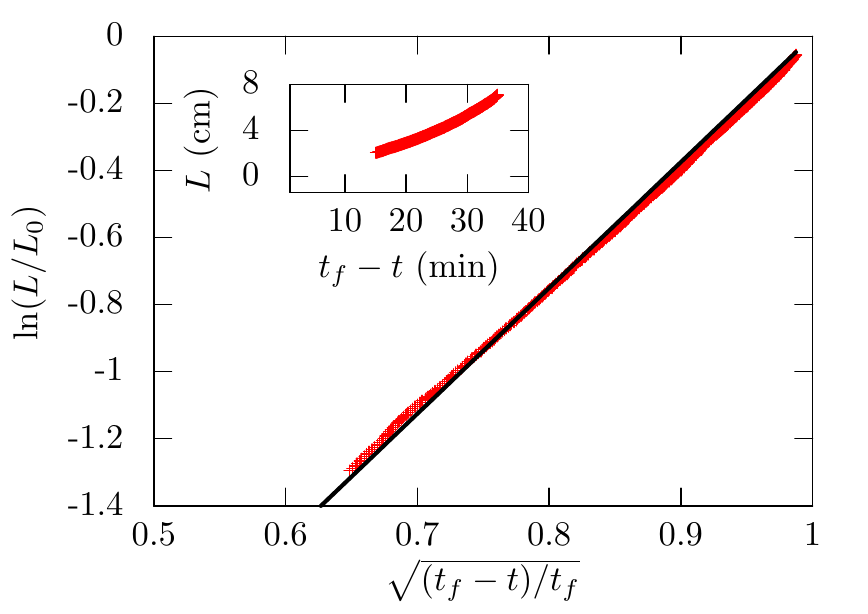}
        \caption{Drying of a long column on two fibers with a tilt angle $\delta=0.45^\circ$.
            The initial length of the liquid (DMTS) column is $L_0=7$ cm and the fiber diameter is $2\,a=0.30$ mm.
            The inset shows the raw data, the length of the liquid column $L$ as a function of time $t_f - t$.
            The main plot represents the data with ${\rm ln}(L/L_0)$ versus $\sqrt{(t_f-t)/t_f}$.
            The black line is the theoretical prediction (equation (\ref{eq:log})) for the drying of long liquid columns between parallel fibers \cite{Duprat2013}.
        }\label{fig:long_col}
    \end{figure}

    We now describe the drying of the liquid in the column morphology, which is observed for small enough tilt angles or volumes of liquid (figure \ref{fig:dyn}).
    In such situations, the Bond number $\textrm{Bo} = \rho_\ell g a^2/\gamma$ that compares gravitational forces and surface tension effects is of the order $10^{-2}$.
    Thus surface tension effects dominate and we shall assume in the following that the liquid maintains a symmetrical shape despite gravitational forces.

    \subsubsection{Small tilt angles and long columns}
    We first consider the case of very long columns obtained for small angles $\delta$ (figure \ref{fig:setup}).
    We define the length $L$ of the column from the node of the fibers to the liquid column extremity as illustrated in the figure \ref{fig:setup}.
    We represent the time evolution of the length $L$ obtained for very small tilt angle, $\delta=0.45^\circ$, and a column of initial length $L_0=7$ cm, in figure \ref{fig:long_col}.
    Note that in the final stages of the drying process, the length of the column cannot be measured near the node.

    In order to model the drying, we assume that the drying of liquid columns between parallel  \cite{Duprat2013} and crossed fibers are similar.
    We assume that the liquid interface has an ellipsoidal shape with a large aspect ratio, \textit{i.e.} $ b \ll L$ where $b =  L \tan(\delta)$ is the distance between the fibers.
    Thus, we set the  lengthscale of the  vapor concentration gradient $\mathcal{L}\approx b \ln (L/b)$.
    As derived in \cite{Duprat2013}, we can show that
    \begin{equation}
        \ln(L/L_0) \propto \sqrt{( t_f - t) / t_f}.\label{eq:log}
    \end{equation}
    Figure \ref{fig:long_col} shows the agreement between the experimental observations and the theoretical prediction given by equation (\ref{eq:log}).
    However, in practice, small tilt angles are relatively rare.
    Indeed, in a randomly oriented network of fibers, the angle distribution can be assumed to be uniform.
    It is therefore important to consider columns formed at relatively larger angles.
    Indeed, as the angle $\delta$ increases, the column length $L$ decreases rapidly  \cite{Sauret2014} and the drying dynamics changes.

    \begin{figure}
        \centering
        \includegraphics[width=0.9\linewidth]{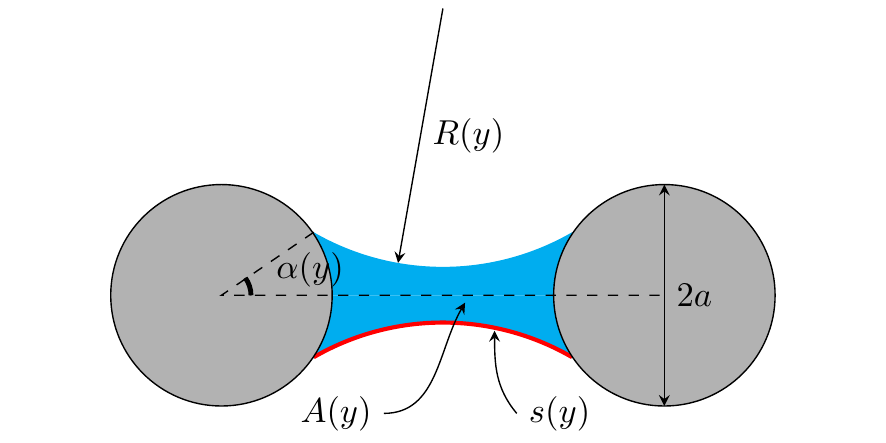}
        \caption{Schematic of the cross section of a liquid in a column morphology between two crossed fibers.
        }\label{fig:theo}
    \end{figure}

    \subsubsection{Modest tilt angles and short columns}\label{sec:columnshort}
    For larger tilt angles $\delta$,  the aspect ratio of the liquid column can no longer be considered large.
    In this case, the time-evolution of the volume of liquid in the column state can be expressed as
    \begin{equation}
        V(t)=2\,\int_{0}^{L(t)}A(y)\,\text{d} y,\label{eq:colvol}
    \end{equation}
    where the factor $2$ corresponds to the two half-columns on each side of the node, and $A(y)$ is the cross-sectional area of the column (figure \ref{fig:theo}).
    Denoting $s(y)$ the arc length of the liquid cross-section, the total interfacial area is given by
    \begin{equation}
        \mathcal{A}(t)=4\,\int_0^{L(t)}s(y)\,\text{d} y=4\,\int_0^{L(t)}\left(\pi-2\,\alpha(y)\right)\,R(y)\,\text{d} y,\label{eq:colarea}
    \end{equation}
    where $\alpha(y)$ and the radius of curvature $R(y)$ of the liquid are defined in  figure \ref{fig:theo} and detailed formulae are given in \cite{Sauret2014}.
    Substituting equations (\ref{eq:colvol}) and (\ref{eq:colarea}) into equation (\ref{eq:master}), we obtain
    \begin{equation}
        \frac{\text{d}}{\text{d} t}\left(\int_{0}^{L(t)}A(y)\,\text{d} y \right)=-\frac{2\,D}{L(t)}\,\int_0^{L(t)}\left(\pi-2\,\alpha(y)\right)\,R(y)\,\text{d} y,
    \end{equation}
    which simplifies to
    \begin{equation}
        \frac{\text{d} L(t)}{\text{d} t}+\frac{2\,D}{L(t)\,A\left[L(t)\right]}\,\int_0^{L(t)}\left(\pi-2\,\alpha(y)\right)\,R(y)\,\text{d} y=0.
        \label{eq:eqdiff_shortcol}
    \end{equation}

\begin{figure}
        \centering
        \includegraphics[width=1\linewidth]{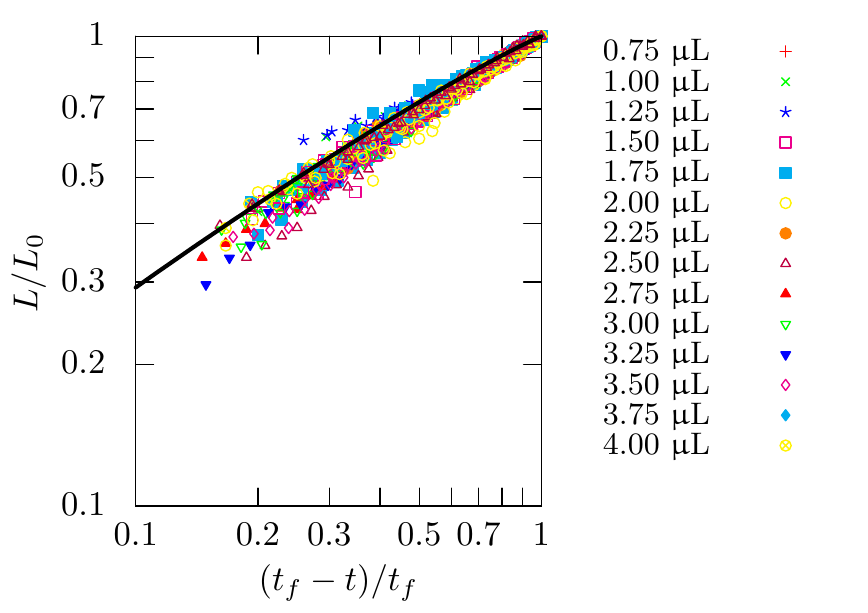}
        \caption{
            Evolution of rescaled length $L/L_0$ for the column state on crossed fibers with rescaled time $(t_f - t)/t_f$ for various initial volumes of OMTS.
            The fiber diameter is $2\,a = 0.28$ mm and the tilt angle is $\delta = 10^\circ$.
            The solid line is the time evolution of the length as predicted by the numerical solution of equation (\ref{eq:eqdiff_shortcol}).
        }\label{fig:short_col}
    \end{figure}

    Experimentally, we choose conditions for which a column is observed at angles $\delta\in[5,16]^\circ$ and we measure the time evolution of the length of the column for different initial volumes.
    These results are rescaled with the initial column length and total evaporation time and reported in figure \ref{fig:short_col}.
    We note that the data all collapse on a master curve and are in good agreement with the numerical solution of equation (\ref{eq:eqdiff_shortcol}).
    In summary, we have succeeded in  describing the drying behavior of long columns (for small tilt angles) and short columns (for larger tilt angles) successfully with our model although the transition between these two regimes may be more complex to capture.
    However, for practical application in an array of fibers, the tilt angle will remain large enough to apply the second situation or small liquid column.

    \subsection{Drying of liquid in the mixed morphology}
\begin{figure}
        \centering
        \includegraphics[width=0.9\linewidth]{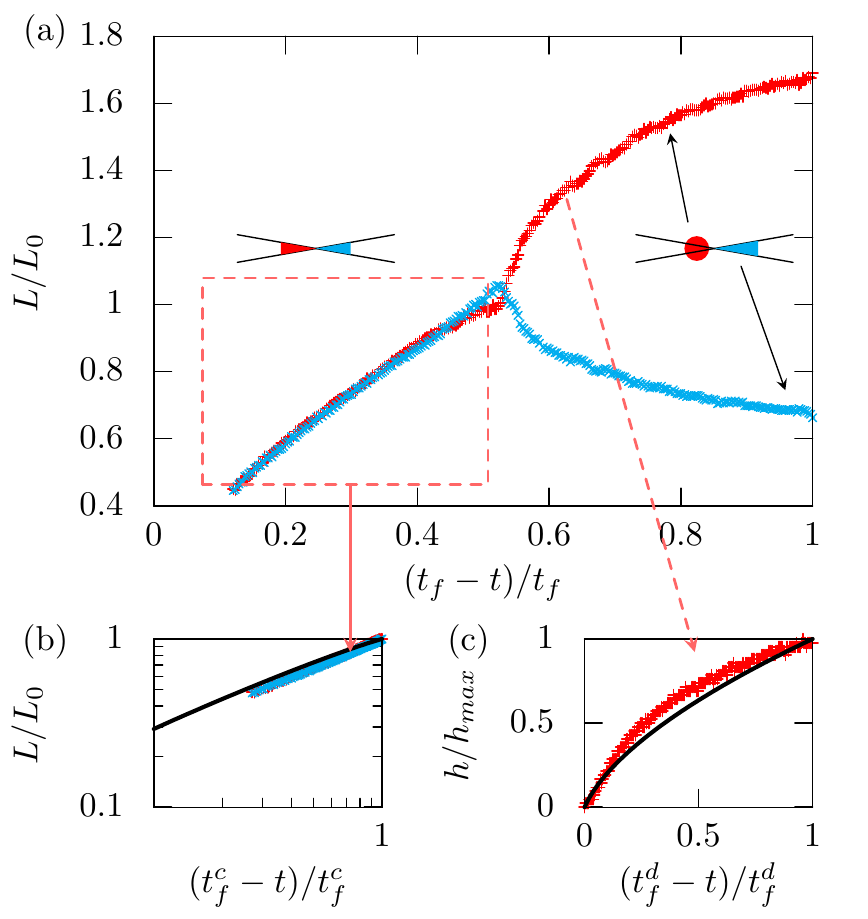}
        \caption{
            Drying dynamics starting from a mixed morphology.
            (a) Evolution of the liquid length on the column and the drop sides with rescaled time $(t_f-t)/t_f$, where $t_f$ is the total drying time.
            The length is normalized by $L_0$, the length of the columns at the morphology transition.
            The red points represent the drop side and the blue points the column side.
            (b) Time evolution of the column lengths in a situation similar to figure \ref{fig:short_col}.
            (c) Time evolution of the drop height as in figure \ref{fig:droplet}(b).
            The fiber diameter is $2\,a=0.30$ mm and the tilt angle is $\delta=20^\circ$.
            Times $t_f^c$ and $t_f^d$ are the drying times for the column and the mixed morphology states, respectively.
        An initial volume of 2 $\mu\ell$ of DMTS is deposited at the node.}\label{fig:mixed}
    \end{figure}

    For intermediate tilt angles between fibers, we observe a mixed morphology (figure \ref{fig:dyn}), which consists of a drop on one side of the node and a liquid column on the other side \cite{Sauret2014}.
    In figure \ref{fig:mixed}(a), we show the time evolution of the liquid length on both sides.
    The dynamics can be split into two regimes.
    Initially, the liquid adopts the mixed morphology state with a large drop in equilibrium with a small column.
    During the drying process,  the volume of the drop decreases while the column elongates slightly to balance the Laplace pressure \cite{Sauret2014}.
    This first regime ends when the two lengths are comparable.
    In the second regime, the drying of two liquid columns (figure \ref{fig:mixed}(b)) follows the dynamics described in section \ref{sec:columnshort}.

    In the first regime, the drop height is measured from the side view.
    The shape is slightly asymmetric due to gravity effects since the Bond number $\textrm{Bo} = \rho_\ell g V^{2/3}/\gamma\sim 10^{-1}$ is not very small, in these experiments.
    Nevertheless, as shown in figure \ref{fig:mixed}(c), the height evolution  behaves similarly as drops on single fibers or at the node of crossed fibers (figure \ref{fig:droplet}).
    Consequently, we show that the evaporation rate in the mixed morphology state is dominated by the drying of the drop.
    When the liquid volume reaches a critical value, the liquid adopts a column shape.
    Figure \ref{fig:mixed}(b) illustrates that the time evolution of the column length is well described by the equation (\ref{eq:eqdiff_shortcol}) for short columns, which is consistent with that of small columns described in the previous section.

    \begin{figure}
        \centering
        \includegraphics[width=0.9\linewidth]{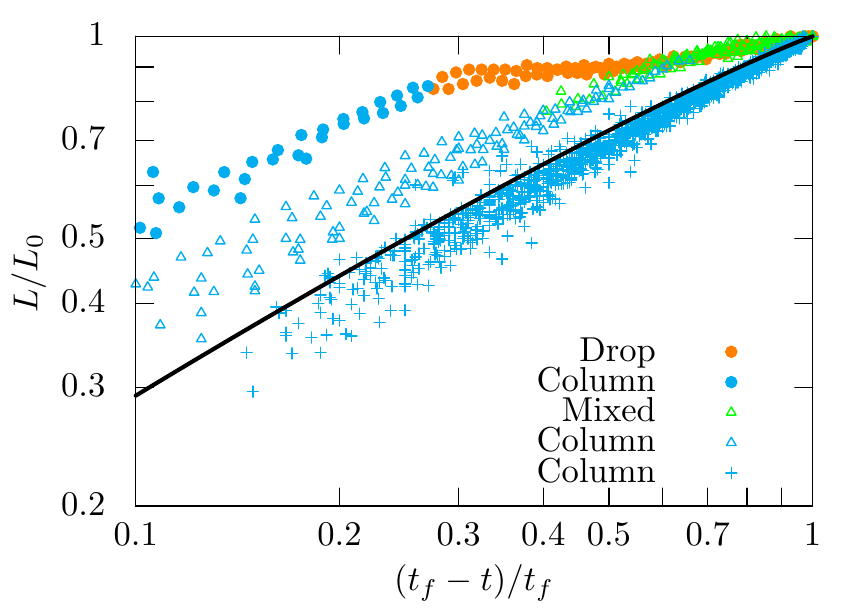}
        \caption{
            Starting from the three possible morphologies, the evolution of the length $L/L_0$ is represented as a function of $(t_f - t)/t_f$.
            The shapes of the symbols correspond to the initial morphology and the color represents the morphology for the given point.
            Data for liquid initially in columns ($+$) is replotted from figure \ref{fig:short_col}.
            The solid line is the evolution of the length with time as predicted by the numerical solution of equation (\ref{eq:eqdiff_shortcol}).
        }\label{fig:allmorphologies}
    \end{figure}

    \subsection{Network of fibers}

    Building on our understanding of the evaporation dynamics in the three morphologies, we compare drying measurements for different tilt angles.
    In figure \ref{fig:allmorphologies}, we present the time evolution of the liquid length for the three different initial morphologies.
    The data illustrate the transition between the liquid morphologies as well as their respective durations of drying.
    We also note that the drying kinetic for liquid columns (blue points in figure \ref{fig:allmorphologies}) follow the same behavior, independent of the initial morphology.
    As discussed previously, due to both the surface area and the characteristic vapor concentration gradient $\mathcal{L}$, the evaporation rate is larger for a column than it is for a drop.

    Now, we apply this finding to the drying of liquid at the nodes of a network of fibers.
    To directly observe the effect of the fiber tilt angles on the drying rate, we make a 2D network of crossed fibers.
    The network is composed of $16$ nodes and each node is about $1$ cm from the neighboring nodes, so that the nodes may be considered independent.
    A volume of $2.7$ $\mu\ell$ of DMTS is deposited on each node.
    The total amount of liquid placed on the network is sufficiently large to measure the mass change with time.
    Thus, the network is placed on a precision scale and the weight is recorded over the drying period.
    The experiment is repeated for different angles between the fibers by shearing the network.
    In tuning the angles, we modify the liquid morphology at the nodes: column, mixed and drop.

    \begin{figure}
        \centering
        \includegraphics[width=0.9\linewidth]{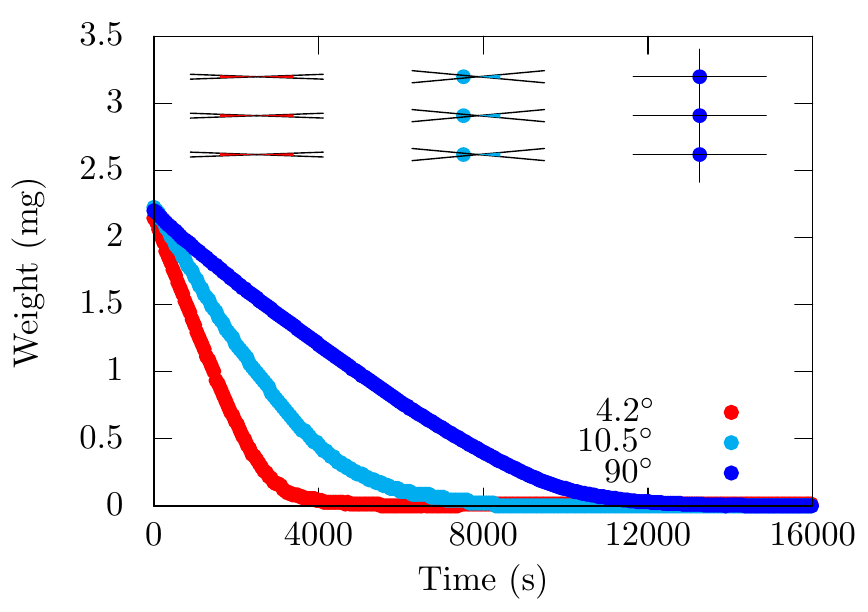}
        \caption{Time evolution of the average liquid weight on each node of a network of crossed fibers making angles $4.2^\circ$, $10.5^\circ$ and $90^\circ$.
            The sketches illustrate the fiber configurations and the initial morphology of the liquid.
            The fiber diameter is $2\,a = 0.30$ mm, the initial liquid (DMTS) volume is $2.7$ $\mu\ell$.
        }\label{fig:array}
    \end{figure}

    The time evolution of the liquid weight on each node is shown in figure \ref{fig:array}.
    The results show that small angles enhance significantly the drying rate, which is consistent with the results obtained with our model system  of two crossed fibers since the liquid tends to be in the column state for which the evaporation rate is increased.
    The total evaporation time for the mixed conformation and the drop are about two and three times the evaporation time of the column, respectively.

    Consequently, for applications in which the drying time needs to be reduced, we show that a network of fibers with small tilt angles performs better.
    Thus, it could be interesting to use fiber networks whose fibers have a preferred orientation.
    Such anisotropy can be, for instance, obtained by shearing an anisotropic material \cite{Borzsonyi2013}.

    \section{Conclusion}
    In this paper, we present an experimental study on the drying of a small liquid volume placed at the node of two crossed fibers with different tilt angles.
    In this configuration, the evaporation of the liquid volume exhibits rich dynamics associated with the different morphologies of the liquid.
    Indeed, the drop, column and mixed morphology evaporate at different rates.
    During the drying process, the liquid can explore several of those morphologies depending upon the initial volume and the angle between the fibers.
    Our experiments show that the time of drying varies significantly between a column from which the liquid evaporates rapidly due to the large surface area and the large aspect ratio, and a drop for which the small surface area leads to low evaporation rates.
    To increase the drying rate, it is thus favorable for the liquid to be in the column morphology.
    This morphology can be achieved by decreasing the tilt angle between fibers as shown here.

    This study provides scaling laws that capture the drying dynamics of the three morphologies under the assumption of a uniform evaporation flux.
    Further studies would be necessary to investigate more accurately the effect of the contact lines on the evaporation flux in this particular geometry \cite{Cazabat2010,Plawsky2008,Wayner1999}.
    We show that the behavior documented on two fibers can also apply to more complex geometries, such as networks of fibers in which nodes are independent.
    The discussion above focuses on drying in a fairly simple geometry, that of two crossed fibers.
    Substrates of higher complexity would render the drying dynamics even more complex with possible interactions between neighboring liquid volumes.

    \section{Acknowledgements}
    We thank C. Duprat, R. A. Register and P. Warren for useful discussions.
    We thank Unilever Research for partial funding of this research.
    F.B. acknowledges that the research leading to these results received funding from the People Programme (Marie Curie Actions) of the European Union's Seventh Framework Programme (FP7/2007-2013) under REA grant agreement 623541.
    \bibliography{article.bib}

\begin{thebibliography}{10}

\bibitem{Shantz1922}
H.L. Shantz.
\newblock {T}ranspiration of plants.
\newblock {\em Botanical Gazette}, 73:239--241, 1922.

\bibitem{Andreas1995}
E.~Andreas.
\newblock {T}he temperature of evaporating sea spray droplets.
\newblock {\em J. Atmos. Sci.}, 52:852--862, 1995.

\bibitem{Reyssat2008}
M.~Reyssat, J.M. Yeomans, and D.~Qu{\'e}r{\'e}.
\newblock {I}mpalement of fakir drops.
\newblock {\em EPL}, 81:26006, 2008.

\bibitem{Anantharaju2009}
N.~Anantharaju, M.~Panchagnula, and S.~Neti.
\newblock {E}vaporating drops on patterned surfaces: {T}ransition from pinned
  to moving triple line.
\newblock {\em Journal of Colloid and Interface Science}, 337:176--182, 2009.

\bibitem{Gelderblom2011}
H.~Gelderblom, A.~Marin, H.~Nair, A.~van Houselt, L.~Lefferts, J.~Snoeijer, and
  D.~Lohse.
\newblock {H}ow water droplets evaporate on a superhydrophobic substrate.
\newblock {\em Phys. Rev. E}, 83:026306, 2011.

\bibitem{Maxwell1877}
J.C. Maxwell.
\newblock {A} treatise on the kinetic theory of gases.
\newblock {\em Nature}, 16:242--246, 1877.

\bibitem{Langmuir1918}
I.~Langmuir.
\newblock {T}he evaporation of small spheres.
\newblock {\em Phys. Rev.}, 12:368--370, 1918.

\bibitem{Cazabat2010}
A.-M. Cazabat and G.~Guena.
\newblock {E}vaporation of macroscopic sessile droplets.
\newblock {\em Soft Matter}, 6:2591--2612, 2010.

\bibitem{Plawsky2008}
Joel~L. Plawsky, Manas Ojha, Arya Chatterjee, and Peter~C. Wayner.
\newblock Review of the effects of surface topography, surface chemistry, and
  fluid physics on evaporation at the contact line.
\newblock {\em Chemical Engineering Communications}, 196(5):658--696, 2008.

\bibitem{Wayner1999}
Peter~C. Wayner.
\newblock Intermolecular forces in phase-change heat transfer: 1998 kern award
  review.
\newblock {\em AIChE Journal}, 45(10):2055--2068, 1999.

\bibitem{Cachile2002}
M.~Cachile, O.~B{\'e}nichou, C.~Poulard, and A.M. Cazabat.
\newblock {E}vaporating droplets.
\newblock {\em Langmuir}, 18:8070--8078, 2002.

\bibitem{Poulard2005}
C.~Poulard, G.~Gu{\'e}na, A.M. Cazabat, A.~Boudaoud, and M.~Ben~Amar.
\newblock {R}escaling the dynamics of evaporating drops.
\newblock {\em Langmuir}, 21:8226--8233, 2005.

\bibitem{Guena2006}
G.~Gu{\'e}na, C.~Poulard, M.~Vou{\'e}, J.D Coninck, and A.M. Cazabat.
\newblock {E}vaporation of sessile liquid droplets.
\newblock {\em Colloids and Surfaces A}, 291:191--196, 2006.

\bibitem{Guena2007}
G.~Gu{\'e}na, C.~Poulard, and A.M. Cazabat.
\newblock {T}he leading edge of evaporating droplets.
\newblock {\em Journal of Colloid and Interface Science}, 312:164--171, 2007.

\bibitem{Berteloot2008}
G.~Berteloot, C.-T. Pham, A.~Daerr, F.~Lequeux, and L.~Limat.
\newblock {E}vaporation-induced flow near a contact line: {C}onsequences on
  coating and contact angle.
\newblock {\em EPL}, 83:14003, 2008.

\bibitem{Pham2010}
C.-T. Pham, G.~Berteloot, F.~Lequeux, and L.~Limat.
\newblock {D}ynamics of complete wetting liquid under evaporation.
\newblock {\em EPL}, 92:54005, 2010.

\bibitem{Lopes2012}
M.~Lopes and E.~Bonaccurso.
\newblock {E}vaporation control of sessile water drops by soft viscoelastic
  surfaces.
\newblock {\em Soft Matter}, 8:7875--7881, 2012.

\bibitem{Clement2004}
F.~Clement and J.~Leng.
\newblock {E}vaporation of liquids and solutions in confined geometry.
\newblock {\em Langmuir}, 20:6538--6541, 2004.

\bibitem{Sauret2014}
A.~Sauret, A.D. Bick, C.~Duprat, and H.A. Stone.
\newblock {W}etting of crossed fibers: {M}ultiple steady states and symmetry
  breaking.
\newblock {\em EPL}, 105:56006, 2014.

\bibitem{Duprat2012}
C.~Duprat, S.~Protiere, A.Y. Beebe, and H.A. Stone.
\newblock {W}etting of flexible fibre arrays.
\newblock {\em Nature}, 482:510--513, 2012.

\bibitem{Quere1988}
D.~Qu{\'e}r{\'e}, J.-M. Di~Meglio, and F.~Brochard-Wyart.
\newblock {W}etting of fibers: theory and experiments.
\newblock {\em Revue Phys. Appl.}, 1988.

\bibitem{Quere1989}
D.~Qu{\'e}r{\'e}, J.-M. Meglio, and F.~Brochard-Wyart.
\newblock {M}aking van der {W}aals films on fibers.
\newblock {\em EPL}, 10:335, 1989.

\bibitem{Quere1999}
D.~Qu{\'e}r{\'e}.
\newblock {F}luid coating on a fiber.
\newblock {\em Annual Review of Fluid Mechanics}, 31:347--384, 1999.

\bibitem{Boulogne2013a}
F.~Boulogne, M.-A. Fardin, S.~Lerouge, F.~Giorgiutti-Dauphin\'{e}, and
  L.~Pauchard.
\newblock Suppression of the {R}ayleigh-{P}lateau instability on a vertical
  fibre coated with wormlike micelle solutions.
\newblock {\em Soft Matter}, 9:7787--7796, 2013.

\bibitem{Boulogne2013b}
F.~Boulogne, L.~Pauchard, and F.~Giorgiutti-Dauphin{\'e}.
\newblock {A}nnular cracks of thin films of colloidal silica particles coating
  a fiber.
\newblock {\em EPL}, 102:39002, 2013.

\bibitem{Rayleigh1878}
L.~Rayleigh.
\newblock {O}n the instability of jets.
\newblock {\em Proc. Lond. Math. Soc.}, s1-10:4--13, 1878.

\bibitem{McHale1999}
G.~McHale, S.M. Rowan, M.I. Newton, and N.A. Kab.
\newblock {E}stimation of contact angles on fibers.
\newblock {\em Journal of Adhesion Science and Technology}, 13:1457--1469,
  1999.

\bibitem{McHale2001}
G.~McHale, M.I. Newton, and B.J. Carroll.
\newblock {T}he shape and stability of small liquid drops on fibers.
\newblock {\em Oil and Gas Science and Technology - Rev. IFP}, 56:47--54, 2001.

\bibitem{Lorenceau2004}
E.~Lorenceau, C.~Clanet, and D.~Qu{\'e}r{\'e}.
\newblock {C}apturing drops with a thin fiber.
\newblock {\em Journal of Colloid and Interface Science}, 279:192--197, 2004.

\bibitem{Princen1970}
H.M Princen.
\newblock {C}apillary phenomena in assemblies of parallel cylinders: {I{I}I}.
  {L}iquid {C}olumns between {H}orizontal {P}arallel {C}ylinders.
\newblock {\em Journal of Colloid and Interface Science}, 34:171--184, 1970.

\bibitem{Protiere2012}
S.~Protiere, C.~Duprat, and H.A. Stone.
\newblock {W}etting on two parallel fibers: drop to column transitions.
\newblock {\em Soft Matter}, 9:271--276, 2012.

\bibitem{Duprat2013}
C.~Duprat, A.~Bick, P.~Warren, and H.A. Stone.
\newblock {E}vaporation of drops on two parallel fibers: {I}nfluence of the
  liquid morphology and fiber elasticity.
\newblock {\em Langmuir}, 29:7857--7863, 2013.

\bibitem{Sultan2005}
E.~Sultan, A.~Boudaoud, and M.~Ben~Amar.
\newblock {E}vaporation of a thin film: diffusion of the vapour and {M}arangoni
  instabilities.
\newblock {\em Journal of Fluid Mechanics}, 543:183--202, 2005.

\bibitem{Maxwell1867}
J.~C. Maxwell.
\newblock {O}n the dynamical theory of gases.
\newblock {\em Philosophical Transactions of the Royal Society of London},
  157:49--88, 1867.

\bibitem{Carroll1976}
B.J. Carroll.
\newblock {T}he accurate measurement of contact angle, phase contact areas,
  drop volume, and {L}aplace excess pressure in drop-on-fiber systems.
\newblock {\em Journal of Colloid and Interface Science}, 57:488--495, 1976.

\bibitem{Borzsonyi2013}
T.~Borzsonyi and R.~Stannarius.
\newblock {G}ranular materials composed of shape-anisotropic grains.
\newblock {\em Soft Matter}, 9:7401--7418, 2013.

\end{thebibliography}
    \bibliographystyle{unsrt}

    \newpage
    ~
    \newpage
    \section{TOC graphic}

        \includegraphics[width=0.9\linewidth]{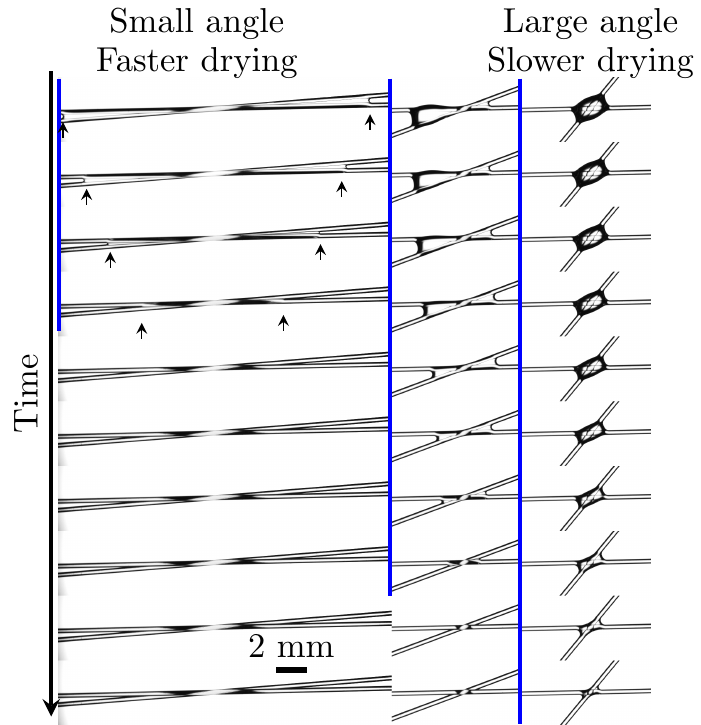}
    \end{document}